\documentclass[cameraready]{Interspeech}

\title{HybridCodec: Fast Dual-Stream, Semantically Enhanced Neural Audio Codec}

\author[affiliation={1}]{Arjun}{Gangwar}
\author[affiliation={1}]{S}{Umesh}

\address{
    $^1$ Indian Institute of Technology, Madras, India
}

\email{arjungangwar@gmail.com, umeshs@ee.iitm.ac.in}


\usepackage{comment}
\usepackage{pifont}
\newcommand{\cmark}{\ding{51}}
\newcommand{\xmark}{\ding{55}}
\usepackage{makecell}
\usepackage{arydshln}
\usepackage{colortbl}
\usepackage[table]{xcolor} 
\usepackage{multirow}

\begin{document}

\maketitle

\begin{abstract}
    The popularity of neural audio codecs as speech tokenizers has surged with the advent of Multimodal Large Language Models. New codec architectures with semantic and acoustic disentanglement have emerged. There are two main approaches to introduce semantic information into codec models: one distills semantic information from SSL representations into the first RVQ layer, while the other maintains separate streams for semantic and acoustic features. We propose HybridCodec, a unified architecture that combines both paradigms. It employs separate semantic and acoustic branches while distilling SSL representations into the semantic stream. This design ensures strong disentanglement without requiring an SSL model during inference. HybridCodec shows superior semantic specialization (RVQ-1) on in-domain test set and competitive reconstruction (RVQ-all). We demonstrate its robustness in out-of-domain and zero-shot cross-lingual settings, achieving a 3x speedup over existing dual-stream models.
\end{abstract}

\section{Introduction}

With the rapid adoption of Multimodal Large Language Models (MLLMs), speech discretization has emerged as a key component for enabling unified modeling across speech and text modalities. Prior works \cite{speech_tokenizer, codec_does_matter, moshi} show that the separation of semantic and acoustic information is crucial; semantic tokens represent linguistic content, while acoustic tokens capture attributes such as speaker identity, prosody, and recording conditions. Effective semantic–acoustic disentanglement allows language models to operate over high-level linguistic content while preserving fine-grained acoustic attributes such as speaker identity and prosody.

Existing speech discretization methods primarily follow two paradigms: semantic distillation-based codecs (e.g., Mimi) \cite{moshi} and dual-stream codecs (e.g., DualCodec) \cite{dualcodec}.

In the dual-stream framework, the model consists of a semantic stream and an acoustic stream. The semantic stream processes representations extracted from a self-supervised learning (SSL) model (e.g., w2v-BERT-2.0) \cite{w2v-bert-2.0} using an autoencoder with a vector quantization (VQ) bottleneck, producing quantized semantic latents. In parallel, the acoustic stream encodes raw audio into latent representations. To encourage disentanglement, the reconstructed semantic latents are subtracted from the acoustic representations before residual vector quantization (RVQ). This encourages the RVQ module to model only residual acoustic information not captured by the semantic stream. During decoding, the semantic and acoustic representations are combined to reconstruct the final waveform.

In contrast, distillation-based approaches adopt a conventional neural codec architecture (e.g., DAC) \cite{DAC} and inject semantic supervision by distilling SSL representations into the first RVQ codebook. This removes the need for a large SSL model at inference time and leads to significantly lower latency.

Although dual-stream codecs demonstrate strong semantic–acoustic disentanglement, often reflected in superior first-codebook (RVQ-1) performance, they incur high inference latency due to the inclusion of large SSL encoders. Distillation-based codecs, on the other hand, achieve fast inference but typically exhibit weaker semantic specialization in the first codebook. Table 1 provides a high-level comparison between these approaches.

In this work, we propose \textbf{HybridCodec}, a unified architecture that combines the strengths of both paradigms. Our model retains the dual-stream structure to explicitly model semantic and acoustic information while leveraging semantic distillation to inject SSL knowledge without requiring a heavyweight SSL model during inference. As a result, HybridCodec achieves inference speeds comparable to distillation-based codecs while preserving the strong semantic–acoustic disentanglement characteristic of dual-stream models. Our main contributions are summarized as follows:
\begin{itemize}
    \item We systematically compare semantic–acoustic disentanglement paradigms under a unified experimental setup, analyzing trade-offs between semantic performance, acoustic quality, and inference speed across in-domain, cross-lingual, and zero-shot settings.
    \item We propose \textbf{HybridCodec}, a dual-stream neural audio codec with semantic distillation that achieves strong RVQ-1 semantic performance while maintaining competitive acoustic quality, low framerate and fast inference.
\end{itemize}

\begin{table}[t]
\caption{High-level overview of codec architectures}
\centering
\resizebox{\columnwidth}{!}{
\begin{tabular}{lcccccc}
\hline
Model &
\makecell{Semantic \\ Enhancement} &
\makecell{Dual \\ Stream} &
Distillation &
\makecell{No SSL model \\ during Inference} &
\makecell{Semantic \\ Disentanglement} &
\makecell{Inference \\ Speed} \\
\hline
DAC           & \xmark & \xmark & \xmark & \cmark & Bad   & Fast \\
DAC (Distill) & \cmark & \xmark & \cmark & \cmark & Good  & Fast \\
DualCodec     & \cmark & \cmark & \xmark & \xmark & Great & Slow  \\
\hdashline
\textbf{HybridCodec}           & \cmark & \cmark & \cmark & \cmark & Great & Fast \\
\hline
\end{tabular}
}
\label{general_overview}
\end{table}

\begin{figure*}[t]
  \centering
  \includegraphics[width=\textwidth]{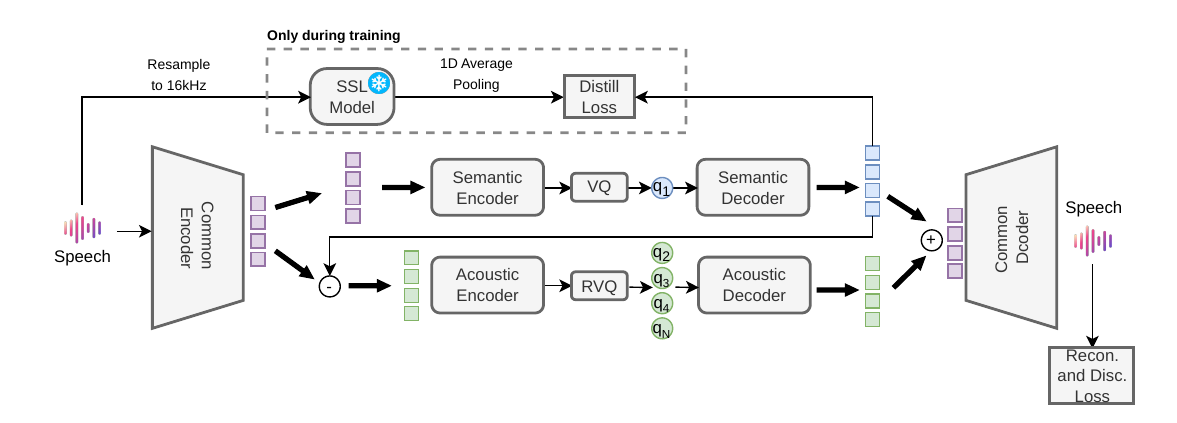}
  \caption{HybridCodec Architecture: Dual-stream with semantic distillation}
  \label{fig:speech_production}
\end{figure*}

\section{Methodology}

As shown in figure 1, we have latents from a common encoder going into two streams - semantic and acoustic streams. Each branch consists of stream specific encoders and decoders. The latents from both the branches are combined and sent to the common decoder which gives out final audio. The output of the semantic decoder distills semantically rich information from SSL representations. This removes the need of carrying a separate SSL model while inference and having two separate streams better disentangle the semantic and acoustic information.


\textbf{Common Encoder and Decoder:} The common encoder and decoder are Convolutional Neural Network (CNN). The common encoder takes $24kHz$ raw waveform as input and applies series strided causal $1D$ convolutions with strides $(4, 5, 6, 8)$. The total downsampling factor for the common encoder is $960$ which when applied on $24kHz$ audio results in latents at $25Hz$ framerate.
The common decoder applies the strided causal $1D$ transpose convolutions to the combined quantized latents with strides $(8, 6, 5, 4)$. This upsamples the combined quantized latents to generate final audio with $24Khz$ sampling rate.


\textbf{Semantic Stream:} The semantic stream consists of a lightweight semantic encoder (SE) and a lightweight semantic decoder (SD) which envelops the vector quantization (VQ) layer. Both the semantic encoder and decoder contain $5$ causal ConvNext blocks\cite{convnext} which maintain the incoming framerates. These blocks provide additional modelling capacity for semantic information.
The semantic encoder takes the latents from the common encoder as input. The output of the common encoder is quantized using the VQ layer giving us RVQ-1 codes. The semantic codebook consists of $16,384$ vectors. The quantized latents are passed to the semantic decoder. The output of the semantic decoder distills semantic information from SSL representations. 
We use $16th$ layer embeddings from w2v-BERT-2.0 \cite{w2v-bert-2.0} as the SSL representations. The model outputs the embeddings at $50Hz$ framerate. So, these embeddings are passed through a $1D$ average pooling layer with stride $2$ to achieve the desired $25Hz$ framerate. We employ simple $L2$-loss between the downsampled SSL embeddings and the semantic decoder output to perform semantic distillation. The SSL model is frozen during training and removed during inference.


\textbf{Acoustic Stream:} The acoustic stream consists of a lightweight acoustic encoder (AE), residual vector quantization (RVQ) layer, and a lightweight acoustic decoder (AD). The acoustic encoder and decoder have the same architecture as semantic encoder and decoder. The output of the semantic decoder is subtracted from the output of the common encoder and passed through the acoustic encoder. This helps with disentanglement of information since the semantic information is not modelled by the acoustic stream. 
The output of the acoustic encoder is discretized by the RVQ layer. The RVQ consists of $N-1$ remaining codebooks. Each acoustic codebook consists of $1024$ vectors. These codes make the RVQ-rest codes. We employ RVQ dropout during training where we use the first $n$ quantizers in each training step where $n \in [0,N-1]$. When $n=0$, we only use the semantic codes.


\textbf{Training Objectives:} Following the framework in DualCodec \cite{dualcodec}, HybridCodec is trained end-to-end. We incorporate a semantic distillation loss alongside the generative adversarial network (GAN) objective from DAC \cite{DAC}. The total objective is composed of the following components:

\begin{itemize}
    \item \textit{Spectrogram Reconstruction Loss ($\mathcal{L}_{spec}$):} A multi-scale Mel-spectrogram loss between the input audio and the reconstructed waveform to ensure acoustic fidelity.
    \item \textit{Semantic Distillation Loss ($\mathcal{L}_{distill}$):} An MSE loss between the semantic decoder's output and the 25Hz downsampled SSL features.
    \item \textit{Quantization Loss ($\mathcal{L}_{q}$):} This includes a \textit{codebook loss} ($\mathcal{L}_{code}$) to update dictionary entries and a \textit{commitment loss} ($\mathcal{L}_{commit}$) to stabilize encoder latents. These are applied to both semantic ($s$) and acoustic ($a$) streams.
    \item \textit{Adversarial Loss ($\mathcal{L}_{adv}$):} Utilizes Multi-Period (MPD) and Multi-Scale STFT (MS-STFTD) discriminators to improve perceptual realism.
    \item \textit{Feature Matching Loss ($\mathcal{L}_{fm}$):} An $L_1$ loss between intermediate discriminator features of real and synthetic samples to stabilize GAN training.
\end{itemize}

The total training objective $\mathcal{L}_{total}$ is defined as:

\begin{equation}
\begin{aligned}
\mathcal{L}_{total} = \lambda_{s} \mathcal{L}_{spec} + \lambda_{d} \mathcal{L}_{distill} + \lambda_{f} \mathcal{L}_{fm} + \lambda_{g} \mathcal{L}_{adv} \\ 
+ \sum_{i \in \{s, a\}} (\lambda_{c} \mathcal{L}_{code}^{i} + \lambda_{m} \mathcal{L}_{commit}^{i})
\end{aligned}
\end{equation}

We set the weight parameters as follows: $\lambda_{s} = 15.0$, $\lambda_{d} = 15.0$, $\lambda_{f} = 2.0$, and $\lambda_{g} = 1.0$. For quantization bottlenecks, the codebook weights $\lambda_{c}$ are set to $1.0$, while the commitment weights $\lambda_{m}$ are $0.25$ for both semantic and acoustic streams.

\begin{table*}[t]
\caption{ Results at 60k updates. All models are trained \textbf{from scratch} on LibriSpeech (24kHz) and have 1 semantic codebook of size 16,384 and 11 acoustic codebooks of size 1024.  }
\centering
\label{table-2-main-60k}
\begin{tabular}{lcccccccc}
\hline
  \multirow{2}{*}{\textbf{Model}} &
  \multicolumn{5}{c}{\textbf{WER ↓}} &
  \multirow{2}{*}{\textbf{SSIM ↑}} &
  \multirow{2}{*}{\textbf{UTMOS ↑}} &
  \multirow{2}{*}{\textbf{PESQ ↑}} \\
\multicolumn{1}{l}{} &
  \textbf{RVQ-1} &
  \textbf{RVQ-1:2} &
  \textbf{RVQ-1:4} &
  \textbf{RVQ-1:8} &
  \textbf{RVQ-1:12} &
   &
   &
   \\ \hline
\multicolumn{9}{c}{Test Clean} \\ \hline
  DAC &
  43.22 &
  14.07 &
  6.22 &
  4.67 &
  4.55 &
  0.6 &
  3.51 &
  2.36 \\
  DAC (Distill) &
  21.54 &
  11.63 &
  6.08 &
  4.52 &
  4.39 &
  0.62 &
  3.53 &
  2.37 \\
  DualCodec &
  18.93 &
  10.11 &
  5.5 &
  4.53 &
  4.8 &
  0.6066 &
  3.4521 &
  2.3073 \\
  \textbf{HybridCodec (HC-SED-AED)} &
  15.36 &
  8.7 &
  5.44 &
  4.96 &
  4.46 &
  0.5777 &
  3.4417 &
  2.2667 \\ \hline
\multicolumn{9}{c}{SeedTTS-en} \\ \hline
  DAC &
  64.85 &
  23.53 &
  7.45 &
  4.08 &
  3.37 &
  0.45 &
  2.95 &
  1.95 \\
  DAC (Distill) &
  39.51 &
  19.78 &
  7.1 &
  3.85 &
  3.45 &
  0.47 &
  3.03 &
  1.94 \\
  DualCodec &
  31.09 &
  12.74 &
  5.06 &
  3.29 &
  3.29 &
  0.4239 &
  2.7029 &
  1.7894 \\
  \textbf{HybridCodec (HC-SED-AED)} &
  30.37 &
  14.2 &
  6.4 &
  4.17 &
  3.96 &
  0.4108 &
  2.9406 &
  1.8808 \\ \hline
\multicolumn{9}{c}{CV-French} \\ \hline
  DAC &
  126.07 &
  92.89 &
  49.83 &
  33.06 &
  31.17 &
  0.3527 &
  1.896 &
  1.804\\
  DAC (Distill) &
  112.56 &
  89.69 &
  51.72 &
  32.35 &
  28.98 &
  0.371 &
  1.946 &
  1.810 \\
  DualCodec &
  103.18 &
  71.05 &
  42.66 &
  29.86 &
  28.41 &
  0.3418 &
  1.788 &
  1.686 \\
  \textbf{HybridCodec (HC-SED-AED)} &
  106.41 &
  77.85 &
  45.52 &
  34.1 &
  32.52 &
  0.333 &
  1.898 &
  1.772 \\ \hline
\end{tabular}
\end{table*}

\section{Experimental Setup}

We train all models on the full 960-hour LibriSpeech corpus \cite{LibriSpeech}, resampling all audio waveforms to 24kHz. During training, audio is randomly cropped into 3-second segments. Optimization is performed for 60k steps on a cluster of 4 NVIDIA RTX A6000 (48GB) GPUs. For evaluation, we utilize the LibriSpeech test-clean sets. To assess out-of-domain (OOD) generalization, we evaluate performance on the SeedTTS-en test set \cite{Anastassiou2024SeedTTSAF}. Furthermore, zero-shot cross-lingual capabilities are tested on 1,000 randomly sampled utterances from the Common Voice French test set \cite{commonvoice-2020}.

We compare our proposed method with retrained baselines for a fair comparison of different architectures. We retrain DualCodec on the 960-hour LibriSpeech data. Additionally, we retrain the DAC and a semantically distilled variant DAC (Distill), using $12$ codebooks with size $N_s = 16$k (semantic) and $N_a = 1$k (acoustic) to match our model's capacity. We have three variant of our model: HC-SE (only semantic encoder), HC-SED (semantic encoder + semantic decoder) and HC-SED-AED (semantic encoder and decoder + acoustic encoder and decoder).

We evaluate intelligibility via Word Error Rate (WER) by transcribing synthesized audio using Whisper v3-large \cite{radford2022whisper}. Perceptual quality is assessed using UTMOS \cite{UTMOS} and PESQ \cite{pesq}. Finally, to evaluate speaker identity preservation, we report Speaker Similarity (SSIM) using cosine similarity between embeddings from a pretrained ECAPA-TDNN \cite{ECAPA-TDNN}.

\section{Results and Discussion}
Table 2 summarizes the performance of all models at 60k updates across in-domain (LibriSpeech Test) and out-of-domain (SeedTTS-en, CV-French) datasets.

\textbf{In-Domain Performance: } On Test Clean, HC-SED-AED achieves the lowest RVQ-1 WER (15.36\%), outperforming both DualCodec (18.93\%) and DAC (Distill) (21.54\%), indicating stronger semantic specialization in the first codebook. As more quantizers are added, performance converges, with HC-SED-AED (4.46\%) remaining competitive with DAC (Distill) (4.39\%) and DualCodec (4.80\%) at RVQ-1:12.

\textbf{Out-of-Domain and Cross-Lingual Generalization: } On SeedTTS-en, HC-SED-AED achieves competitive RVQ-1 WER (30.37\%), closely matching DualCodec (31.09\%) and outperforming DAC (Distill) (39.51\%). Final RVQ performance remains comparable across models. In the challenging CV-French zero-shot setting, HC-SED-AED (106.41\%) improves over DAC (Distill) (112.56\%) and approaches DualCodec (103.18\%) in RVQ-1 WER. Performance differences narrow at higher RVQ levels, suggesting robust acoustic modeling.

Overall, HC-SED-AED consistently improves semantic (RVQ-1) performance over both DualCodec and DAC (Distill) on in-domain data, while maintaining competitive reconstruction quality and strong generalization, all without requiring an SSL model during inference. We can further tradeoff slight RVQ-1 performance for better acoustic performance by making some architectural choices as discussed in section 4.3.

\begin{table*}[t]
\caption{Effect of extended training till 300k updates. The open-source checkpoints are taken from the original repositories.}
\centering
\label{results-at-300k}
\begin{tabular}{lrrrrrccc}
\hline
\multicolumn{1}{c}{\multirow{2}{*}{\textbf{Model}}} &
  \multicolumn{5}{c}{\textbf{WER ↓}} &
  \multirow{2}{*}{\textbf{SSIM ↑}} &
  \multirow{2}{*}{\textbf{UTMOS ↑}} &
  \multirow{2}{*}{\textbf{PESQ ↑}} \\
\multicolumn{1}{c}{} &
  \multicolumn{1}{c}{\textbf{RVQ-1}} &
  \multicolumn{1}{c}{\textbf{RVQ-1:2}} &
  \multicolumn{1}{c}{\textbf{RVQ-1:4}} &
  \multicolumn{1}{c}{\textbf{RVQ-1:8}} &
  \multicolumn{1}{c}{\textbf{RVQ-1:12}} &
   &
   &
   \\ \hline
\multicolumn{9}{c}{\textbf{Test Clean}}                                                            \\ \hline
Mimi (open-source)    & 35.77 &  8.58         & 6.32          & 4.15          & 4.14          &  0.8374 &  3.7745 &  2.589 \\
DualCodec (open-source)    & 17.23 &  14.04         & 8.22          & 5.02          & 6.32          &  0.92 &  3.771 &  2.833 \\ \hdashline
DAC (Distill)    & 18.01 & 10.55         & 5.34          & 4.82          & 4.14          & 0.776 & 3.926 & 2.834 \\
DualCodec      & 12.75 & 7.49          & 4.61          & 3.93          & 3.87          & 0.823 & 3.95 & 2.954 \\
\textbf{HC-SED-AED} & 12.96 & 7.32 & 4.4  & 3.83 & 3.63 & 0.816 & 3.99 & 2.894 \\ \hline
\multicolumn{9}{c}{\textbf{SeedTTS-en}}                                                            \\ \hline
Mimi (open-source)    & 41.33 &  10.82         & 6.12          & 3.48          & 3.06          &  0.7994 &  3.4008 &  2.3993 \\
DualCodec (open-source)    & 10.46 &  8.66         & 4.11          & 2.42          & 2.39          &  0.853 &  3.435 &  2.621 \\ \hdashline
DAC (Distill)    & 30.32 & 15.73         & 4.68          & 3             & 3.05          & 0.628 & 3.483 & 2.315 \\
DualCodec      & 17.7  & 8.56          & 4.18          & 2.85          & 2.69          & 0.717 & 3.525 & 2.489 \\
\textbf{HC-SED-AED} & 20.02 & 8.94          & 3.52          & 2.94          & 2.91          & 0.687 & 3.556 & 2.352 \\ \hline
\multicolumn{9}{c}{\textbf{CV-French}}                                                             \\ \hline
Mimi (open-source)    & 99.39 &  65.4         & 43.13          & 26.29          & 19.27          &  0.742 & 2.182  & 2.243  \\
DualCodec (open-source)    & 54.8 &  47.29         & 26.36          & 16.83          & 14.12          & 0.808  & 2.234  &  2.544 \\ \hdashline
DAC (Distill)    & 109.4 & 83.97         & 38.43         & 22.44         & 19.72         & 0.561 & 2.223 & 2.188 \\
DualCodec      & 84.33 & 56.92         & 28.69         & 18.81         & 17.11         & 0.657 & 2.263 & 2.321 \\
\textbf{HC-SED-AED} & 91.75 & 57.98         & 27.65         & 18.63         & 17.25         & 0.632 & 2.281 & 2.182 \\ \hline
\end{tabular}
\end{table*}

\subsection{Effect of extended training and Comparision with open-source models}

Table~\ref{results-at-300k} analyzes the impact of extended training upto 300k updates. Across all datasets, increasing training steps substantially improves WER for all models, particularly at lower quantization levels. On Test Clean, HC-SED-AED reduces RVQ-1 WER from 15.36\% (60k) to 12.96\% (300k), while also achieving the best performance at higher RVQ levels (e.g., 4.46\%  to 3.63\% at RVQ-1:12), outperforming both DualCodec and DAC (Distill).

For out-of-domain datasets, additional training further improves generalization. On SeedTTS-en, RVQ-rest WER consistently decreases, indicating better acoustic reconstruction, while maintaining competitive semantic (RVQ-1) performance. On the challenging CV-French zero-shot setting, extended training significantly reduces WER across all RVQ levels, narrowing the gap with DualCodec and demonstrating improved cross-lingual robustness.

Overall, scaling training from 60k to 300k updates leads to consistent gains in both semantic specialization and reconstruction fidelity. Notably, HC-SED-AED benefits more prominently at deeper RVQ levels, suggesting that longer training particularly strengthens acoustic modeling while preserving semantic disentanglement.

\subsection{Real-Time Factor (RTF) and Throughput}
We evaluate inference efficiency on a single NVIDIA RTX A6000 (48GB) GPU, as summarized in Table~\ref{rtf_analysis}. To ensure measurement accuracy, we performed five warm-up iterations followed by three independent runs with GPU synchronization to account for asynchronous kernel execution. The reported values are the average of these runs. As shown in Table~\ref{rtf_analysis}, HC-SED-AED achieves an RTF of 0.014, representing a 3$\times$ speedup over DualCodec (RTF 0.042). This gain is primarily due to our hybrid approach, which leverages semantic distillation to eliminate the high computational overhead of the 600M-parameter SSL encoder required by DualCodec at inference. While the single-stream DAC (Distill) remains the fastest, HC-SED-AED maintains a high throughput of 348.75 samples/sec. This demonstrates that our dual-stream architecture provides superior semantic disentanglement with minimal latency, offering a practical balance for real-time applications.

\begin{table}[t]
\caption{RTF and Throughput of different models}
\label{rtf_analysis}
\centering
\resizebox{\columnwidth}{!}{
\begin{tabular}{lcccc}
\hline
Model & \#Params & RTF ↓ & \makecell{Throughput \\ RTF ↓} & Samples/sec ↑ \\
\hline
DAC (Distill) & 98.48M & 0.005 & 0.001 & 934 \\
DualCodec & 102.29M + 600M & 0.042 & 0.009 & 114.963 \\
\textbf{HC-SED-AED} & 159.62M & 0.014 & 0.003 & 348.747 \\
\hline
\end{tabular}
}
\end{table}

\subsection{Ablation Study: Model Architecture}

Table~\ref{table5-ablation} illustrates the performance trade-offs across our three architectural variants: HC-SE, HC-SED, and the full HC-SED-AED.

\textbf{Semantic-Acoustic Trade-off:} The full HC-SED-AED architecture achieves the best semantic specialization, reaching a best-in-class RVQ-1 WER of 15.36\% on Test Clean. This confirms that explicitly modeling residuals via an acoustic stream is critical for disentanglement. However, this comes with a slight reduction in acoustic fidelity (PESQ/SSIM) due to the stricter information bottleneck.

\textbf{Balancing Quality and Intelligibility:} If higher reconstruction quality is prioritized over absolute semantic purity, HC-SED serves as an optimal middle ground. It improves WER over the HC-SE while maintaining better acoustic metrics (PESQ 2.34 vs.\ 2.26) than the full HC-SED-AED variant.

\textbf{Robustness:} Both dual-stream variants (HC-SED and HC-SED-AED) significantly outperform the HC-SE in zero-shot (CV-French) and OOD (SeedTTS-en) settings.

While HC-SED-AED is superior for speech tokenization which requires high semantic disentanglement, HC-SED offers a compelling alternative for applications where acoustic fidelity is as vital as intelligibility.

\begin{table}[t]
\centering
\footnotesize
\setlength{\tabcolsep}{4pt}   
\caption{Ablation on architectural choices (60k updates)}
\label{table5-ablation}

\begin{tabular}{lccccc}
\hline
\multirow{2}{*}{\textbf{Model}} &
\multicolumn{2}{c}{\textbf{WER ↓}} &
\multirow{2}{*}{\textbf{SSIM ↑}} &
\multirow{2}{*}{\textbf{UTMOS ↑}} &
\multirow{2}{*}{\textbf{PESQ ↑}} \\

& \textbf{RVQ-1} & \textbf{RVQ-all} & & & \\
\hline

\multicolumn{6}{c}{\textbf{Test Clean}} \\
\hline
HC-SE       & 19.70 & 4.45 & 0.6125 & 3.4360 & 2.3178 \\
HC-SED       & 18.38 & 4.51 & 0.6150 & 3.4597 &2.3470\\
HC-SED-AED & 15.36 & 4.46  & 0.5777 &  3.4417 & 2.2667 \\


\hline
\multicolumn{6}{c}{\textbf{SeedTTS-en}} \\
\hline
HC-SE       &34.41& 3.44 & 0.4514 & 2.9887 & 1.9392 \\
HC-SED       & 31.15 & 3.44 & 0.4482 & 3.0189 & 1.9408 \\
HC-SED-AED & 30.37 & 3.96 & 0.4108 & 2.9406 & 1.8808\\

\hline
\multicolumn{6}{c}{\textbf{CV-French}} \\
\hline
HC-SE       & 117.72 & 30.47 & 0.383 & 1.939 & 1.84 \\
HC-SED      & 111.38 & 28.56 & 0.38 & 1.96 & 1.846 \\
HC-SED-AED & 106.41 & 32.52 & 0.333 & 1.898 & 1.772 \\

\hline
\end{tabular}
\end{table}

\section{Conclusion}

We presented HybridCodec, a dual-stream neural audio codec that bridges the gap between semantic disentanglement and inference efficiency. By utilizing a dual-stream architecture with SSL distillation, HybridCodec eliminates the need for heavyweight SSL encoders at inference. We show that HybridCodec achieves superior semantic specialization, yielding the lowest RVQ-1 WER (15.36\%) on LibriSpeech, while delivering a 3$\times$ speedup over DualCodec. Our evaluations across out-of-domain and cross-lingual tasks confirm that this hybrid approach maintains high-fidelity reconstruction and robust generalization. Overall, our findings show that combining dual-stream disentanglement with distillation provides a practical and effective solution for speech tokenization.

\section{Generative AI Use Disclosure }
We used ChatGPT 5.2 and Gemini 3 for minor language editing and phrasing improvements. All scientific content, experimental design, and conclusions are solely the work of the authors.

\bibliographystyle{IEEEtran}
\bibliography{mybib}

\end{document}